\documentclass[12pt]{article}
\usepackage{placeins}
\usepackage{mathtools}
\usepackage{flafter}
\usepackage{natbib}
\usepackage{hyperref}
\usepackage{graphicx}
\usepackage{booktabs}
\usepackage{geometry}
\usepackage{amssymb}
\newcommand{\dd}{\mathrm{d}}
\newcommand{\pp}{\partial}
\newcommand{\Bo}{\mathit{Bo}}

\title{Curvature-corrected sloshing spectra for cylindrical tanks in microgravity}
\author{Gianni Cassoni\thanks{Corresponding author: \texttt{gianni.cassoni@uniroma3.it}}\\[4pt]
\small Department of Civil, Computer Science and Aeronautical Technology Engineering,\\
\small Roma Tre University, Via Vito Volterra 62, Rome, Italy}
\date{}

\begin{document}
\maketitle

\begin{abstract}
In microgravity, a partially filled cylindrical tank is generally bounded by a curved equilibrium meniscus rather than by an almost flat free surface. This modifies both the bulk liquid inertia and the capillary restoring force, so flat-interface sloshing frequencies can become inaccurate even in the linear regime. This effect matters once the Bond number is of order unity or smaller, precisely the regime relevant to capillarity-dominated propellant management.
This study revisits the classical cylindrical curved-meniscus eigenvalue problem for capillary--gravity sloshing about axisymmetric Young--Laplace equilibria. A semi-analytical boundary-operator formulation is derived that preserves the cylindrical Bessel structure and recovers the flat-interface limit exactly. Its main advantage lies in treating the bulk Dirichlet--Neumann operator and the linearised curvature operator as distinct components, thereby making the physical origin of curvature-induced frequency shifts explicit.
The results show that equilibrium curvature couples radial modes and alters the low-order spectrum once $\Bo\lesssim 1$. Concave menisci lower the fundamental frequency, whereas convex menisci raise it while often lowering higher branches. The asymmetry between wetting and non-wetting configurations is found to be predominantly kinetic, being carried mainly by the Dirichlet--Neumann operator rather than by the capillary term. Curved menisci should therefore be treated as part of the leading-order model of cylindrical microgravity sloshing, not as a secondary correction, if reduced-order predictions are to capture the relevant dynamical scales for spacecraft applications.
\end{abstract}

\section{Introduction}
\label{sec:intro}

Free-surface oscillations in partially filled containers are a classical problem of fluid mechanics and remain central to propellant storage, transfer and management in low-gravity environments \citep{simonini2024}. Recent microgravity experiments and direct numerical simulations in spherical tanks have further shown that capillarity-dominated interface dynamics can generate measurable force, torque and bubble-oscillation signatures during manoeuvres \citep{dalmon2019}. For a partially filled cylindrical tank, the relevant control parameter is the Bond number $\Bo=\rho g R^{2}/\sigma$, where $\rho$ is the liquid density, $g$ the gravitational acceleration, $R$ the tank radius and $\sigma$ the surface tension. When $\Bo\gg 1$, gravity maintains an almost flat interface and the classical cylindrical Bessel--Fourier spectrum is an accurate approximation. When $\Bo\lesssim 1$, by contrast, capillarity and wetting enter at leading order. The static contact angle $\theta_c$, measured through the liquid at the wall, selects a curved equilibrium meniscus, and this curvature modifies both the liquid inertia and the restoring force. Only in the special case $\theta_c=\pi/2$ does the equilibrium interface remain flat and the low-gravity cylindrical problem remain separable \citep{myshkis1987}; when $\theta_c\neq\pi/2$, the flat-interface theory no longer provides the appropriate reference problem.

The appropriate comparison class is therefore not the broad open-surface gravity--capillary literature, but the confined-wave tradition in which edge constraints, contact-line physics and meniscus geometry enter at leading order. This viewpoint emerged in studies of closed basins and edge-constrained capillary--gravity waves \citep{miles1976,benjaminscott1979}, and was then carried into cylindrical configurations through analyses of resonant and forced modal interactions \citep{miles1984a,miles1984b,funakoshi1988}, wetting and non-wetting boundary conditions \citep{cocciaro1991,cocciaro1993}, and contact-line-induced spectral effects \citep{miles1991,kidambi2009}. More recent contributions in {\it Journal of Fluid Mechanics} have continued this confined-wave programme in cylindrical settings through studies of contact-angle hysteresis, orbital forcing and nearly brimful meniscus waves \citep{viola2018,bongarzone2022a,bongarzone2022b}. These works show that confined capillary--gravity dynamics remains an active topic, but they are concerned primarily with damping, hysteresis and forced or parametric response rather than with the low-Bond eigenvalue problem about prescribed Young--Laplace equilibria.

For low-gravity cylindrical tanks, the benchmark literature developed along a separate but complementary line. Classical engineering sloshing theory supplied flat-interface modal descriptions and equivalent mechanical models for partially filled containers \citep{abramson1966,dodge2000}. A capillary-hydrostatic viewpoint was then developed for weak-$g$ environments, where interface location and shape are controlled by surface tension and wetting once body forces cease to dominate \citep{reynolds1964}. Early simulated-low-gravity experiments documented the corresponding departure from terrestrial sloshing behaviour \citep{dodgegarza1967}. In parallel, Soviet/Russian analyses developed the mathematical structure of the curved-meniscus problem by incorporating surface forces into oscillation theory for axisymmetric vessels and by treating capillary equilibrium, stability and small oscillations in weak force fields \citep{moiseev1965,temkin1972,slobozhanin1973,slobozhanin1974,kopachevskii1973}. This line culminated in the systematic cylindrical Ritz treatment compiled by \citet{myshkis1987}. Operator-theoretic developments later recast sloshing and related free-surface problems in Hilbert-space form \citep{kopachevskii2001,kopachevskii2003,kuznetsov2002,kozlov2004,kuznetsov2021}. What remains less transparent in the classical cylindrical setting, however, is how much of the curvature-induced frequency shift is kinetic, through the modification of the bulk Laplace problem, and how much is capillary, through the linearised curvature of the equilibrium meniscus.

This separation becomes natural in the Zakharov--Craig--Sulem formulation \citep{zakharov1968,craig1993,zakharov2002}, where irrotational free-surface motion is expressed in terms of the surface elevation and the trace of the velocity potential, while the bulk problem is encoded by the Dirichlet--Neumann operator. In that representation, the linearised dynamics split naturally into a kinetic contribution, carried by the bulk harmonic extension, and a restoring contribution, carried by gravity and the linearised curvature operator. The present study exploits this structure to develop a semi-analytical boundary-operator formulation for the classical cylindrical curved-meniscus eigenproblem about axisymmetric Young--Laplace equilibria. The formulation preserves the cylindrical Bessel basis, recovers the flat-interface limit exactly, and allows direct comparison with the cylindrical Ritz benchmark values reported by \citet{kopachevskii1973} and reproduced in the monograph of \citet{myshkis1987}. Its specific contribution is to keep the bulk inertial and capillary restoring parts explicit, so that the physical origin of curvature-induced frequency shifts can be resolved rather than inferred only from the combined spectrum. In particular, it becomes possible to identify which part of the shift, and of the wetting/non-wetting asymmetry, is carried by the Dirichlet--Neumann operator and which part by the linearised curvature operator. In this sense, the classical low-Bond cylindrical eigenproblem is recast in the modern boundary-operator language used in capillary--gravity wave analysis \citep{craignicholls2000,alazardbaldi2015,bertifeolafranzoi2021}, while retaining the natural Bessel structure of the cylindrical tank.
\section{Hamiltonian formulation}
\label{sec:hamiltonian}

Consider an irrotational, incompressible fluid of density $\rho$ occupying a rigid cylindrical tank of radius $R$. The free surface is assumed to remain a graph

\[
  z=\eta(t,r,\theta)
\]

over the horizontal disk

\[
  \mathcal{S}:=\{(r,\theta): 0\le r\le R,\; 0\le \theta<2\pi\},
\]

so the fluid domain is

\begin{equation}
  \label{eq:domain}
  \mathcal{D}_\eta
  := \bigl\{(r,\theta,z):
       (r,\theta)\in\mathcal{S},\;
       0\le z\le \eta(t,r,\theta)\bigr\}.
\end{equation}

The flow is irrotational and incompressible, so the bulk velocity field can be written as

\[
  \boldsymbol{u}=\nabla\Phi,
  \qquad
  \Delta\Phi=0 \quad \text{in } \mathcal{D}_\eta,
\]

and the liquid volume is prescribed:

\[
  \int_{\mathcal{S}} \eta\,\dd A = V_0 = \pi R^{2} h,
\]

where $h$ is the mean fill depth and $\dd A = r\,\dd r\,\dd\theta$.

Following \citet{zakharov1968,craig1993}, the dynamics are written in the conjugate surface variables $\bigl(\eta,\psi\bigr)$, where

\[
  \psi(t,r,\theta):=\Phi\bigl(t,r,\theta,\eta(t,r,\theta)\bigr)
\]

is the trace of the velocity potential on the free surface. Thus $\boldsymbol{u}$ denotes the bulk velocity field, whereas $\psi$ denotes only the surface trace of $\Phi$ on the free surface. On the manifold of prescribed volume, the evolution equations take the Hamiltonian form

\begin{equation}
  \label{eq:ham-sys}
  \pp_t \begin{pmatrix} \eta \\ \psi \end{pmatrix}
  = \mathcal{J}
    \begin{pmatrix} \delta_\eta \mathcal{H} \\[2pt]
                     \delta_\psi \mathcal{H} \end{pmatrix},
  \qquad
  \mathcal{J}
  = \begin{pmatrix} 0 & 1 \\ -1 & 0 \end{pmatrix},
\end{equation}

where $\delta_\eta \mathcal{H}$ and $\delta_\psi \mathcal{H}$ denote the variational derivatives of $\mathcal{H}$ with respect to $\eta$ and $\psi$, respectively,

with Hamiltonian (energy per unit density)

\begin{equation}
  \label{eq:hamiltonian}
  \mathcal{H}(\eta,\psi)
  = \frac{1}{2}\int_{\mathcal{S}}
      \psi\,G(\eta)\,\psi\,\dd A
  + \frac{g}{2}\int_{\mathcal{S}}
      \eta^{2}\,\dd A
  + \kappa\int_{\mathcal{S}}
      \Bigl(\sqrt{1+|\nabla_s\eta|^{2}}-1\Bigr)\,\dd A
\end{equation}

where $\kappa:=\sigma/\rho$ and $\nabla_s=(\pp_r,\,r^{-1}\pp_\theta)$ is the horizontal gradient and $G(\eta)$ is the Dirichlet--Neumann operator. In the present formulation, \eqref{eq:hamiltonian} is used only as a formal starting point for the derivation of the linearised bulk kinetic and interior capillary contributions. No explicit wall-energy or contact-line term is included at the Hamiltonian level. Instead, the analysis is carried out about a prescribed axisymmetric Young--Laplace equilibrium meniscus $H_0$, and the reduced dynamics are obtained by linearisation on the admissible perturbation class associated with that equilibrium.
\subsection{The Dirichlet--Neumann operator}
\label{ssec:DN-def}

The only non-local ingredient in \eqref{eq:hamiltonian} is the Dirichlet--Neumann operator $G(\eta)$, which encodes the bulk Laplace problem at the free surface. Let

\[
  \Gamma_\eta:=\{(r,\theta,z): z=\eta(t,r,\theta)\}
\]

denote the free surface, and let $\Phi$ be the harmonic extension into $\mathcal{D}_\eta$ of a prescribed surface potential $\psi$, that is,

\[
  \Delta\Phi=0 \quad \text{in } \mathcal{D}_\eta,
  \qquad
  \Phi|_{\Gamma_\eta}=\psi,
\]

with rigid-wall boundary conditions

\[
  \pp_z\Phi|_{z=0}=0,
  \qquad
  \pp_r\Phi|_{r=R}=0.
\]

The Dirichlet--Neumann operator is then defined by

\begin{equation}
  \label{eq:DN-def}
  G(\eta)\,\psi
  := \bigl(\pp_z\Phi-\nabla_s\eta\cdot\nabla_s\Phi\bigr)\big|_{z=\eta}.
\end{equation}

Equivalently,

\[
  G(\eta)\,\psi
  = \sqrt{1+|\nabla_s\eta|^{2}}\,
    \pp_n\Phi\big|_{\Gamma_\eta},
\]

where $\pp_n$ denotes differentiation along the outward upward unit normal to $\Gamma_\eta$. Thus $G(\eta)$ maps the surface potential $\psi$ to the corresponding weighted normal derivative of its harmonic extension.

This definition is naturally associated with the Hamiltonian structure through the Dirichlet energy. Let $\psi_1$ and $\psi_2$ be two traces on $\Gamma_\eta$, and let $\Phi_1$ and $\Phi_2$ denote their harmonic extensions into $\mathcal{D}_\eta$, satisfying the same rigid-wall boundary conditions. Green's identity then gives

\begin{equation}
  \label{eq:DN-green}
  \int_{\mathcal{D}_\eta}
  \left(
    \Phi_2\,\Delta\Phi_1
    + \nabla\Phi_1\cdot\nabla\Phi_2
  \right)\,\dd V
  = \int_{\Gamma_\eta} (\pp_n\Phi_1)\,\Phi_2\,\dd\Gamma
  = \int_{\mathcal{S}} G(\eta)\,\psi_1\,\psi_2\,\dd A
\end{equation}

where $\dd V$ and $\dd\Gamma$ denote the volume and surface elements, $\Delta\Phi_1=0$, and the last equality uses the Craig--Sulem convention that absorbs the surface Jacobian $\sqrt{1+|\nabla_s\eta|^{2}}$ into $G(\eta)$. Hence $G(\eta)$ is self-adjoint and non-negative with respect to the flat $L^{2}(\mathcal{S})$ inner product, and its kernel consists of constants.

In the flat cylindrical limit, $G$ is diagonal in the Bessel--Fourier basis. For the axisymmetric equilibrium meniscus considered below, azimuthal Fourier modes proportional to $e^{im\theta}$ decouple. Accordingly, a single azimuthal wavenumber $m$ is fixed in what follows, and the operator is subsequently evaluated at the equilibrium meniscus $H_0$.

\subsection{Static equilibrium and linearisation}
\label{sec:equilibrium}
\label{ssec:linearise}

The reduced formulation is built about a prescribed axisymmetric equilibrium meniscus

\[
  z=H_0(r)
\]

of fixed volume. In the computations below, the equilibrium family is parameterised by the wall slope at the sidewall, equivalently by the static contact angle, but this parameter does not enter through the Hamiltonian formulation itself. In the presence of surface tension, $H_0$ satisfies the Young--Laplace equation

\begin{equation}
  \label{eq:young-laplace}
  \frac{\sigma}{r}\,\frac{\dd}{\dd r}
    \left(\frac{r\,H_0'}{\sqrt{1+H_0'^{\,2}}}\right)
  = \rho g\,H_0 - \mu,
\end{equation}

where $\mu$ is a Lagrange multiplier associated with the prescribed liquid volume $V_0$. The boundary conditions are

\begin{equation}
  \label{eq:bc-meniscus}
  H_0'(0)=0,
  \qquad
  H_0'(R)=\cot\theta_c.
\end{equation}

\begin{figure}[htbp]
  \centerline{\includegraphics[width=\textwidth,height=0.42\textheight,keepaspectratio]{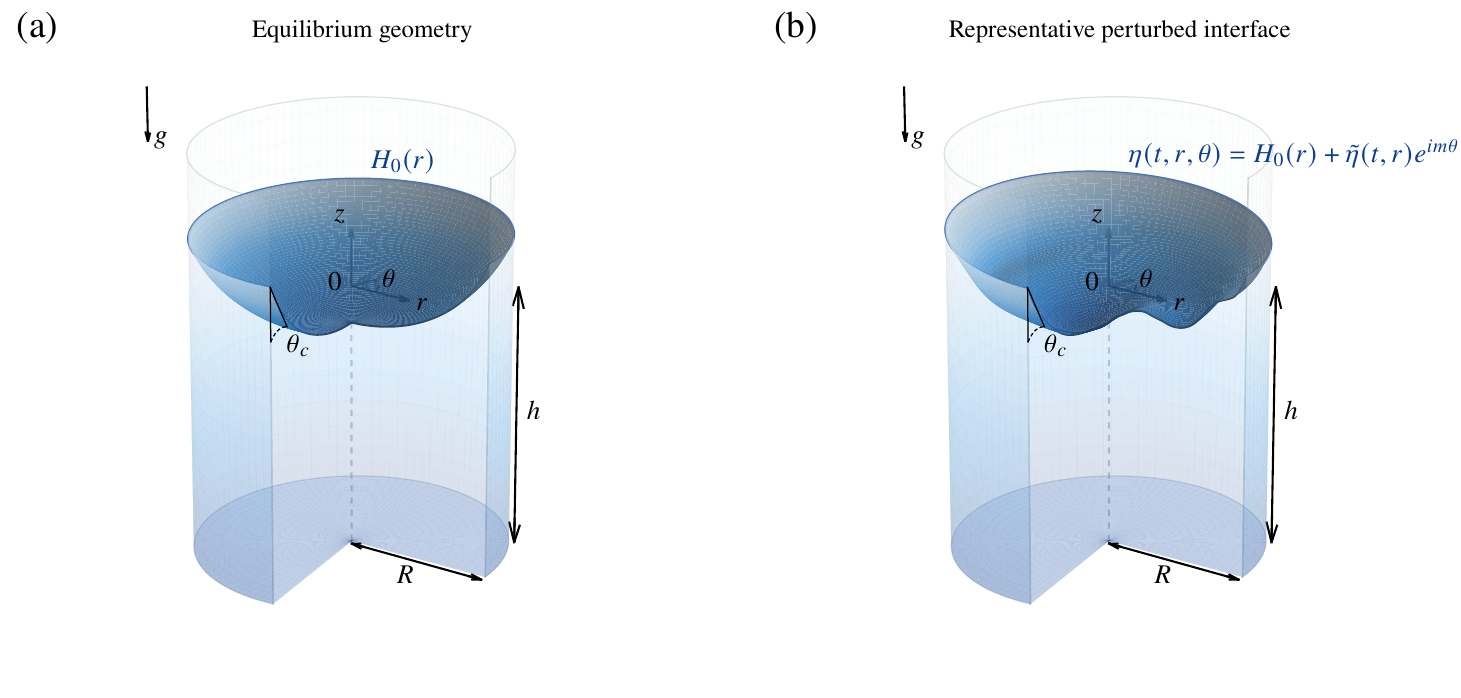}}
  \caption{Geometry sketch for the cylindrical free-surface problem. Panel (a) shows the axisymmetric equilibrium meniscus $z=H_0(r)$ obtained from the Young--Laplace problem \eqref{eq:young-laplace}--\eqref{eq:bc-meniscus}. Panel (b) shows a representative perturbed interface in a fixed azimuthal sector,
  $\eta(t,r,\theta)=H_0(r)+\tilde\eta(t,r)e^{im\theta}$.}
  \label{fig:geometry-sketch}
\end{figure}

A schematic of the equilibrium geometry and of a representative perturbed interface is shown in figure~\ref{fig:geometry-sketch}.

The first condition enforces regularity at the symmetry axis; the second prescribes the static contact angle at the wall. For wetting liquids ($\theta_c<90^\circ$) the meniscus rises at the wall, while for non-wetting liquids ($\theta_c>90^\circ$) it dips.

The present formulation requires the equilibrium interface to remain a graph over the horizontal plane. This holds for all contact angles $\theta_c>0$ at sufficiently large $\Bo$, but for very small $\theta_c$ and $\Bo$ the meniscus may approach a nearly closed surface, in which case the graph parametrisation, the vertical diffeomorphism and the Bessel--Fourier basis cease to apply. The analysis is therefore restricted to menisci that intersect the sidewall at a finite contact angle with moderate slope.
Equation \eqref{eq:young-laplace} is solved numerically by introducing the slope-angle variable

\[
  \beta(r):=\arctan(H_0'(r))
\]

following \citet{concus1968}. The Young--Laplace equation then becomes a first-order system in $(H_0,\beta)$ parametrised by $r$, subject to $\beta(0)=0$ and $\beta(R)=\pi/2-\theta_c$. For each trial $\mu$, the system is integrated from $r=0$ to $r=R$ with a high-order Runge--Kutta scheme, and $\mu$ is adjusted by Brent's method until the contact-angle condition at the wall is satisfied. The volume constraint

\[
  2\pi\int_0^R H_0\,r\,\dd r = V_0
\]

is then enforced by a uniform vertical shift. This shooting procedure retains the full nonlinear curvature. For more general geometries, Young--Laplace equilibria may instead be obtained by constrained capillary-energy minimisation methods such as the Surface Evolver \citep{brakke1992}; see also \citep{brakke1978} for the broader mean-curvature-flow framework.

The reduced surface dynamics are obtained by linearising about the equilibrium state $(\eta,\psi)=(H_0,0)$, where $H_0(r)$ is the static meniscus and $\psi=0$ denotes the surface potential at rest. For a fixed azimuthal Fourier mode $e^{im\theta}$, write

\[
  \eta(t,r,\theta)=H_0(r)+\tilde\eta(t,r)e^{im\theta},
  \qquad
  \psi(t,r,\theta)=\psi(t,r)e^{im\theta},
\]

and suppress the common factor $e^{im\theta}$ in what follows. Admissible perturbations are taken to preserve the wall slope of the prescribed equilibrium to first order, so that

\[
  \tilde\eta'(R)=0,
\]

together with the linearised volume constraint

\[
  \int_{\mathcal{S}} \tilde\eta\,\dd A = 0,
\]

which is automatic for $m\neq 0$.

At equilibrium $\psi=0$, and since the kinetic term in \eqref{eq:hamiltonian} is quadratic in $\psi$, its shape derivative does not contribute at first order. Linearising \eqref{eq:ham-sys} about $(H_0,0)$ therefore gives

\begin{subequations}
\label{eq:lin-ham}
\begin{align}
  \pp_t\tilde\eta &= G[H_0]\,\psi,
    \label{eq:lin-kin}\\
  \pp_t\psi &= -g\,\tilde\eta
    + \kappa\,\delta\mathcal{C}[H_0]\,\tilde\eta,
    \label{eq:lin-dyn}
\end{align}
\end{subequations}

where $G[H_0]$ is the Dirichlet--Neumann operator evaluated at the equilibrium meniscus and $\delta\mathcal{C}[H_0]$ is the linearisation about $H_0$ of the graph-curvature operator

\[
  \mathcal{C}[\eta]
  := \nabla_s\cdot
  \left(
    \frac{\nabla_s\eta}{\sqrt{1+|\nabla_s\eta|^{2}}}
  \right).
\]

For the $m$th azimuthal mode, this linearised curvature operator is

\begin{equation}
  \label{eq:jacobi}
  \delta\mathcal{C}[H_0]\,\tilde\eta
  = \frac{1}{r}\frac{\dd}{\dd r}\left(
      \frac{r\,\tilde\eta'}
           {\bigl(1+H_0'^{\,2}\bigr)^{3/2}}
    \right)
  - \frac{m^{2}}{r^{2}}\,
      \frac{\tilde\eta}
           {\bigl(1+H_0'^{\,2}\bigr)^{1/2}}.
\end{equation}

This is the Jacobi operator associated with the equilibrium graph $H_0$.

On the admissible class of perturbations, $\delta\mathcal{C}[H_0]$ is negative-semidefinite. Multiplying \eqref{eq:jacobi} by $\tilde\eta\,r$ and integrating by parts yields

\begin{equation}
  \label{eq:neg-def}
  \int_0^R \tilde\eta\,\delta\mathcal{C}[H_0]\,\tilde\eta\;r\,\dd r
  = -\int_0^R \left[
      \frac{(\tilde\eta')^{2}}
           {(1+H_0'^{\,2})^{3/2}}
    + \frac{m^{2}}{r^{2}}\,
      \frac{\tilde\eta^{2}}
           {(1+H_0'^{\,2})^{1/2}}
    \right] r\,\dd r
  \le 0,
\end{equation}

where the contribution at $r=0$ vanishes by axis regularity and the boundary term at $r=R$ vanishes because $\tilde\eta'(R)=0$. For $m=0$, the constant null mode is excluded by the linearised volume constraint. Hence both terms on the right-hand side of \eqref{eq:lin-dyn} are restoring: the gravity term contributes $-g\,\tilde\eta$, and the capillary term contributes $\kappa\,\delta\mathcal{C}[H_0]\tilde\eta$ with $\delta\mathcal{C}[H_0]\le 0$.

\section{Discrete surface operators}
\label{sec:galerkin}

With the equilibrium meniscus $H_0$ fixed, the linearised problem for a given azimuthal wavenumber $m$ reduces to the construction of finite-dimensional approximations of the operators appearing in \eqref{eq:lin-ham}. The Dirichlet--Neumann operator $G[H_0]$ is realised through harmonic lifts in the fixed equilibrium domain $\mathcal{D}$,

\[
  \mathcal{D}
  := \bigl\{(r,\theta,z):
       0\le r\le R,\;
       0\le \theta<2\pi,\;
       0\le z\le H_0(r)\bigr\},
\]

and, after restriction to a single azimuthal Fourier mode $e^{im\theta}$, through the corresponding meridional domain

\[
  \mathcal{D}_{rz}
  := \bigl\{(r,z):
       0\le r\le R,\;
       0\le z\le H_0(r)\bigr\}.
\]

The capillary and mass operators, by contrast, act directly on the free surface. For a chosen truncation order $N$, the discretisation therefore employs two spaces with distinct roles: the surface space $X_N^m$ for the free-surface variables and the auxiliary bulk harmonic space $Y_N^m$ for their harmonic extensions. The vertical diffeomorphism introduced above serves only to motivate the choice of bulk trial functions; all matrices are assembled directly in the physical equilibrium domain. Since a single azimuthal Fourier mode $e^{im\theta}$ is fixed throughout, the dependence on $\theta$ is suppressed in both the surface and bulk variables. Surface pairings are therefore written in the reduced inner product

\[
  \langle f,g\rangle_r := \int_0^R f(r)\,g(r)\,r\,\dd r,
\]

which is the $\theta$-reduced form of the flat $L^2(\mathcal S)$ inner product.
\subsection{Diffeomorphism and trial-space motivation}
\label{ssec:diffeo}

The bulk harmonic trial space is motivated by the vertical diffeomorphism

\[
  T\colon (r,z)\longmapsto (r,\zeta),
  \qquad
  \zeta=\frac{z}{H_0(r)},
\]

which maps the meridional equilibrium domain $\mathcal D_{rz}$ to the flat reference domain

\[
  \hat{\mathcal D}=\{0\le r\le R,\; 0\le \zeta\le 1\}.
\]

For a fixed azimuthal wavenumber $m$, write the potential as
\[
  \Phi(r,\theta,z)=\Phi(r,z)e^{im\theta},
\]
with the same symbol used for the reduced meridional amplitude by abuse of notation. Since
\[
  \Delta
  = \pp_{rr}+\frac{1}{r}\pp_r+\frac{1}{r^2}\pp_{\theta\theta}+\pp_{zz},
\]
the Laplace equation $\Delta\Phi=0$ reduces to
\[
  \Phi_{rr}+\frac{1}{r}\Phi_r-\frac{m^2}{r^2}\Phi+\Phi_{zz}=0
  \qquad \text{in } \mathcal D_{rz}.
\]

together with the rigid boundary conditions

\[
  \pp_z\Phi|_{z=0}=0,
  \qquad
  \pp_r\Phi|_{r=R}=0.
\]

Upon writing

\[
  \phi(r,\zeta):=\Phi\bigl(r,\zeta H_0(r)\bigr),
  \qquad
  J(r):=H_0(r),
  \qquad
  \alpha(r):=H_0'(r),
\]

the Laplace equation becomes

\begin{equation}
  \label{eq:laplace-mapped}
  \frac{1+\alpha^2\zeta^2}{J^2}\,\pp_{\zeta\zeta}\phi
  + \pp_{rr}\phi + \frac{1}{r}\pp_r\phi
  - \frac{2\alpha\zeta}{J}\,\pp_{r\zeta}\phi
  + \frac{\zeta}{J}\left(\frac{2\alpha^2}{J}-\alpha'-\frac{\alpha}{r}\right)\pp_\zeta\phi
  - \frac{m^2}{r^2}\,\phi = 0.
\end{equation}

The meniscus shape therefore enters only through the $r$-dependent coefficients $J$, $\alpha$ and $\alpha'$, while the reference domain remains fixed. In the flat limit $H_0\equiv h$, one has $J\equiv h$ and $\alpha\equiv 0$, so \eqref{eq:laplace-mapped} reduces to the standard separable cylindrical Laplace equation. The corresponding non-constant separated modes are

\[
  \cosh(\epsilon_{ms}\zeta h/R)\,J_m(\epsilon_{ms}r/R),
\]

where $J_m$ denotes the Bessel function of the first kind of order $m$, and $\epsilon_{ms}$ is the $s$th positive zero of $J_m'$. This observation motivates the normalised cosh-weighted Bessel fields

\begin{equation}
  \label{eq:cosh-basis}
  \Phi_s(r,z)
  = \frac{\cosh(\epsilon_{ms}z/R)}
         {\cosh(\epsilon_{ms}\bar h_s/R)}
    J_m\left(\frac{\epsilon_{ms}}{R}r\right),
  \qquad s=1,\ldots,N,
\end{equation}

where $\bar h_s$ is a mode-dependent reference depth introduced only to improve conditioning. These fields are exact harmonic functions in the physical equilibrium domain $\mathcal D$ and satisfy the rigid-wall and bottom boundary conditions by construction. Curvature destroys separability at the equilibrium meniscus, but not harmonicity in the bulk. Indeed,

\[
  \Phi_s(r,H_0(r))
  = \frac{\cosh(\epsilon_{ms}H_0(r)/R)}
         {\cosh(\epsilon_{ms}\bar h_s/R)}
    J_m\left(\frac{\epsilon_{ms}}{R}r\right),
\]

so if $H_0$ is not constant, the prefactor depends on $r$ and a single Bessel mode no longer remains isolated on the surface. This is the origin of the radial-mode coupling that must be resolved by the discrete trace map. The diffeomorphism therefore identifies the curved-meniscus problem as a variable-coefficient perturbation of the flat cylindrical one and thereby motivates the choice of bulk trial space. The harmonic extensions and the resulting surface operators are assembled directly in the physical domain $\mathcal D$, rather than by numerical solution of the mapped equation. The discrete spaces used in the Galerkin construction are therefore introduced as a surface trial space for the free-surface variables and an auxiliary bulk harmonic space for their harmonic extensions,

\[
  X_N^m := \operatorname{span}\{\varphi_n\}_{n=1}^N,
  \qquad
  Y_N^m := \operatorname{span}\{\Phi_s\}_{s=1}^N.
\]

The surface basis functions are chosen as

\begin{equation}
  \label{eq:basis}
  \varphi_n(r)=J_m(\epsilon_{mn}r/R),
  \qquad n=1,\ldots,N,
\end{equation}

where $\epsilon_{mn}$ is the $n$th positive zero of $J_m'$. These functions satisfy the wall condition

\[
  \pp_r\varphi_n|_{r=R}=0
\]

and are orthogonal on $[0,R]$ with weight $r$:

\begin{equation}
  \label{eq:ortho}
  \int_0^R \varphi_p(r)\,\varphi_q(r)\,r\,\dd r
  = \frac{R^{2}}{2}\left(1-\frac{m^{2}}{\epsilon_{mq}^{2}}\right)
    J_m^{2}(\epsilon_{mq})\,\delta_{pq}
  =: \mathcal{N}_q\,\delta_{pq}.
\end{equation}

The bulk basis functions are the normalised cosh-weighted Bessel fields $\Phi_s$ defined in \eqref{eq:cosh-basis}. Thus $\{\varphi_n\}_{n=1}^N$ spans the surface trial space, whereas $\{\Phi_s\}_{s=1}^N$ spans the auxiliary bulk harmonic space.

\subsection{Harmonic extension and trace matching}
\label{ssec:surface-coupling}

For each surface basis function $\varphi_q\in X_N^m$, let $\Phi^{(q)}\in Y_N^m$ denote its discrete harmonic lift, that is, the element of the bulk trial space whose trace on the equilibrium meniscus $z=H_0(r)$ reproduces the prescribed surface datum $\varphi_q$ in the Galerkin sense. Since $Y_N^m$ is spanned by the bulk harmonic fields $\{\Phi_s\}_{s=1}^N$, $\Phi^{(q)}$ is sought in the form

\begin{equation}
  \label{eq:expansion}
  \Phi^{(q)}(r,z)
  = \sum_{s=1}^{N} c_s^{(q)}\,\Phi_s(r,z).
\end{equation}

Every finite linear combination in \eqref{eq:expansion} is harmonic in the fluid domain and satisfies the rigid-wall and bottom boundary conditions by construction. The remaining requirement is the Dirichlet trace condition at the equilibrium meniscus. Because $\Phi^{(q)}$ is constrained to the finite-dimensional bulk space $Y_N^m$, this condition is imposed in projected form: the trace mismatch is required to be orthogonal to the surface space $X_N^m$,

\begin{equation}
  \label{eq:surface-bc}
  \left\langle \Phi^{(q)}(r,H_0(r)),\varphi_p\right\rangle_r
  =
  \langle \varphi_q,\varphi_p\rangle_r,
  \qquad p=1,\ldots,N.
\end{equation}

Substitution of \eqref{eq:expansion} into \eqref{eq:surface-bc} yields the linear system

\[
  \sum_{s=1}^{N} A_{ps}\,c_s^{(q)}
  = \mathcal{N}_q\,\delta_{pq},
  \qquad p=1,\ldots,N,
\]

where the trace-coupling matrix is

\begin{equation}
  \label{eq:coupling-matrix}
  A_{ps}
  = \int_0^R
    \frac{\cosh(\epsilon_{ms}H_0(r)/R)}
         {\cosh(\epsilon_{ms}\bar h_s/R)}
    \,\varphi_p(r)\,\varphi_s(r)\,r\,\dd r.
\end{equation}

The matrix $\mathbf A$ is purely kinematic: it represents the projected trace map from the bulk harmonic space $Y_N^m$ to the surface space $X_N^m$. It is neither a bulk Laplacian matrix nor the Dirichlet--Neumann matrix itself. Its role is solely to describe how the trace of each bulk basis function decomposes in the surface basis.

The reference depths $\bar h_s$ enter only through this conditioning rescaling. In the flat case $H_0\equiv h$, each bulk mode remains proportional to a single surface Bessel mode, so $\mathbf A$ is diagonal. For a curved meniscus, by contrast, the trace of a bulk mode on $z=H_0(r)$ acquires an $r$-dependent prefactor, so a single bulk mode projects onto several surface modes. The matrix $\mathbf A$ is therefore dense, and the coefficients are computed by truncated singular-value decomposition.

\subsection{Dirichlet--Neumann matrix}
\label{ssec:dirichlet-energy}

For $\varphi_p,\varphi_q\in X_N^m$, define

\begin{equation}
  \label{eq:Gpq}
  G_{pq}
  := \langle \varphi_p,\,G[H_0]\varphi_q\rangle_r.
\end{equation}

Let $\Phi^{(q)}$ and $\Phi^{(p)}$ denote the corresponding harmonic lifts. Application of \eqref{eq:DN-green} at $\eta=H_0$ gives

\begin{equation}
  \label{eq:G-energy}
  G_{pq}
  = \int_{\mathcal D_{rz}}
    \left[
      \pp_r\Phi^{(p)}\,\pp_r\Phi^{(q)}
      + \pp_z\Phi^{(p)}\,\pp_z\Phi^{(q)}
      + \frac{m^2}{r^2}\,\Phi^{(p)}\,\Phi^{(q)}
    \right] r\,\dd z\,\dd r,
\end{equation}

The discrete Dirichlet--Neumann matrix is therefore assembled from the bulk Dirichlet-energy pairing of the harmonic lifts, rather than by explicit evaluation of the curved-surface normal derivative. The matrix $\mathbf A$ determines the extension coefficients through projected trace matching, whereas $\mathbf G$ records the resulting kinetic bilinear form on the surface space.

Using the discrete lift expansion \eqref{eq:expansion} and introducing $k_s:=\epsilon_{ms}/R$ substitution into \eqref{eq:G-energy} yields

\begin{equation}
  \label{eq:G-bilinear}
  G_{pq}
  = \sum_{s,t=1}^{N}
    c_s^{(p)}\,K_{st}\,c_t^{(q)},
\end{equation}

where \(K_{st}\) denotes the reduced bulk Dirichlet-energy Gram matrix of the basis \(\{\Phi_s\}_{s=1}^N\).

Evaluation of the bulk Dirichlet-energy integral gives

\begin{equation}
  \label{eq:K-kernel}
  \begin{split}
    K_{st}
    = \int_0^R \Bigl[
        \varphi_s'(r)\,\varphi_t'(r)\,I_{st}^{cc}(r)
        + \varphi_s(r)\,\varphi_t(r)
          \Bigl(
            k_s k_t\,I_{st}^{ss}(r)
            + \frac{m^2}{r^2}\,I_{st}^{cc}(r)
          \Bigr)
      \Bigr] r\,\dd r
  \end{split}
\end{equation}

where

\begin{align}
  I_{st}^{cc}(r)
  &= \int_0^{H_0(r)}
    \frac{\cosh(k_s z)}{\cosh(k_s\bar h_s)}
    \frac{\cosh(k_t z)}{\cosh(k_t\bar h_t)}\,\dd z,
    \label{eq:Icc}\\
  I_{st}^{ss}(r)
  &= \int_0^{H_0(r)}
    \frac{\sinh(k_s z)}{\cosh(k_s\bar h_s)}
    \frac{\sinh(k_t z)}{\cosh(k_t\bar h_t)}\,\dd z.
    \label{eq:Iss}
\end{align}

Thus $\mathbf K$ is the Dirichlet-energy Gram matrix of the bulk basis, whereas the coefficient vectors $\boldsymbol c^{(q)}$ transfer this bulk information back to the surface basis. The $z$-integrals in \eqref{eq:Icc}--\eqref{eq:Iss} are explicit, and the remaining radial integrals are evaluated by Gauss--Legendre quadrature. Since \eqref{eq:G-energy} is a Dirichlet-energy identity, $\mathbf G$ is symmetric and positive semidefinite by construction.
The present discrete formulation is closely related to the cylindrical variational approximation of Myshkis et al. \citep{myshkis1987}, which employs the same class of cosh-weighted Bessel trial fields for the bulk potential. It differs mainly in that the harmonic extension, the projected trace map and the resulting surface kinetic operator are introduced separately, so that the roles of $\mathbf A$ and $\mathbf G$ remain explicit.

\subsection{Capillary and mass matrices}
\label{ssec:capillary}

The remaining matrices act directly on the surface space $X_N^m$ and do not require any bulk extension. Projection of $-\delta\mathcal{C}[H_0]$ onto the surface basis gives

\begin{equation}
  \label{eq:Lpq}
  L_{pq}
  = \int_0^R \left[
      \frac{\varphi_p'\,\varphi_q'}
           {(1+H_0'^{\,2})^{3/2}}
    + \frac{m^{2}}{r^{2}}\,
      \frac{\varphi_p\,\varphi_q}
           {(1+H_0'^{\,2})^{1/2}}
    \right] r\,\dd r.
\end{equation}

By \eqref{eq:neg-def}, $-\delta\mathcal{C}[H_0]$ is non-negative on the admissible class, so its Galerkin matrix $\mathbf L$ is symmetric and positive semidefinite. For $H_0\equiv h$,

\[
  L_{pq}^{\mathrm{flat}}
  = (\epsilon_{mp}/R)^{2}\,\mathcal{N}_p\,\delta_{pq}.
\]

For a curved meniscus, the $r$-dependent weights break Bessel orthogonality and generate off-diagonal coupling. The mass matrix is the reduced $L^2$ Gram matrix of the surface basis:

\begin{equation}
  \label{eq:Mpq}
  M_{pq}
  = \langle \varphi_p,\varphi_q\rangle_r
  = \mathcal{N}_q\,\delta_{pq}.
\end{equation}

Because the basis is orthogonal, $\mathbf M$ is diagonal and coincides with the diagonal matrix on the right-hand side of the trace-matching system.

\subsection{Generalised eigenvalue problem}
\label{ssec:eigen}

The matrices $\mathbf G$, $\mathbf L$ and $\mathbf M$ now close the Galerkin discretisation of the linearised problem. The remaining step is to express the surface dynamics in the finite-dimensional space $X_N^m$ and reduce them to an algebraic eigenvalue problem for the modal frequencies. With

\[
  \tilde\eta=\sum_q a_q\,\varphi_q,
  \qquad
  \psi=\sum_q b_q\,\varphi_q,
\]

projection of \eqref{eq:lin-ham} onto $X_N^m$ gives

\begin{subequations}
\label{eq:galerkin-sys}
\begin{align}
  -i\omega\,M_{pq}\,a_q &= G_{pq}\,b_q,
    \label{eq:gal-kin}\\
  -i\omega\,M_{pq}\,b_q &=
    -\bigl(g\,M_{pq}+\kappa\,L_{pq}\bigr)\,a_q.
    \label{eq:gal-dyn}
\end{align}
\end{subequations}

Here $\mathbf L$ represents $-\delta\mathcal{C}[H_0]$ on the surface space, so the restoring matrix is $g\,\mathbf M+\kappa\,\mathbf L$. The coefficient vectors $(a_q)$ and $(b_q)$ therefore describe the free-surface displacement and surface potential in the Bessel basis. Eliminating $\boldsymbol{b}$ gives

\begin{equation}
  \label{eq:eliminate-discrete}
  \omega^{2}\,\mathbf M\,\mathbf G^{-1}\mathbf M\,\boldsymbol{a}
  = \bigl(g\,\mathbf M+\kappa\,\mathbf L\bigr)\,\boldsymbol{a}.
\end{equation}

\section{Results}
\label{sec:results}


\subsection{Flat-surface limit and verification}
\label{ssec:flat}

For $\theta_c=90^\circ$ the equilibrium meniscus is flat, $H_0\equiv h$, and the discrete problem is diagonal in the Bessel basis. In particular,

\[
  G_{pq}
  = \frac{\epsilon_{mp}}{R}\tanh\!\left(\frac{\epsilon_{mp}h}{R}\right)\mathcal{N}_p\,\delta_{pq},
  \qquad
  L_{pq}
  = \left(\frac{\epsilon_{mp}}{R}\right)^2 \mathcal{N}_p\,\delta_{pq},
  \qquad
  M_{pq}
  = \mathcal{N}_p\,\delta_{pq}.
\]

The generalised eigenproblem \eqref{eq:eliminate-discrete} therefore decouples mode by mode and yields the classical cylindrical capillary--gravity dispersion relation

\begin{equation}
  \label{eq:disp}
  \omega_{mn}^{2}
  =
  \left(
    g+\frac{\sigma}{\rho}\frac{\epsilon_{mn}^{2}}{R^{2}}
  \right)
  \frac{\epsilon_{mn}}{R}
  \tanh\!\left(\frac{\epsilon_{mn}h}{R}\right).
\end{equation}

Table~\ref{tab:flat} shows that, for the test case $(\theta_c,\Bo,m,h/R,N)=(90^\circ,0,1,1,8)$, the computed nondimensional eigenvalues

\[
  \lambda_n:=\omega_{1n}^2\rho R^3/\sigma
\]

agree with the analytical values from \eqref{eq:disp} to machine precision.

\begin{table}
 \begin{center}
  \begin{tabular}{cccc}
    \toprule
    Mode $n$ & $\lambda_n$ (computed) & $\lambda_n$ (analytical) & Rel. error \\
    \midrule
    1 & 5.932  & 5.932  & $<10^{-13}$ \\
    2 & 151.83 & 151.83 & $<10^{-13}$ \\
    3 & 613.0  & 613.0  & $<10^{-13}$ \\
    4 & 1571   & 1571   & $<10^{-13}$ \\
    5 & 3204   & 3204   & $<10^{-13}$ \\
    6 & 5694   & 5694   & $<10^{-13}$ \\
    \bottomrule
  \end{tabular}
  \caption{Flat-surface verification ($\theta_c=90^\circ$, $\Bo=0$, $m=1$, $h/R=1$, $N=8$): computed and analytical eigenfrequencies from \eqref{eq:disp}. The nondimensional eigenvalue is $\lambda=\omega_{1n}^2\rho R^3/\sigma$.}
  \label{tab:flat}
  \end{center}
\end{table}

\subsection{Comparison with benchmark values}

A second benchmark is provided by the cylindrical weak-gravity results reported by \citet{kopachevskii1973} and reproduced in the monograph of \citet{myshkis1987} for pure capillary oscillations at $\Bo=0$, $h/R=1$ and $m=1$. The comparison is made in terms of the fundamental nondimensional eigenvalue

\[
  \lambda_{1n}:=\omega_{1n}^{2}\rho R^{3}/\sigma,
  \qquad n=1,2.
\]

The values reported in table~\ref{tab:myshkis} show that the present formulation reproduces this benchmark very well for the fundamental branch and satisfactorily for the second branch. For $\lambda_{11}$, the agreement is within $0.6\%$ for $30^\circ\le\theta_c\le165^\circ$ and within $0.1\%$ for $60^\circ\le\theta_c\le120^\circ$; the largest discrepancy, $5.7\times10^{-3}$, occurs at $\theta_c=15^\circ$. For $\lambda_{12}$, the relative error stays below $0.8\%$ for $30^\circ\le\theta_c\le120^\circ$, is about $1.9\%$ at $15^\circ$, and rises to a few per cent for the steepest non-wetting menisci. This deterioration is consistent with the stronger ill-conditioning of the trace-coupling matrix as the wall slope increases. Since the equilibrium meniscus itself is computed from the full nonlinear Young--Laplace problem, the remaining discrepancies are more plausibly attributable to Galerkin truncation, quadrature and conditioning in the projected trace solve than to the equilibrium description.

Broader benchmark data for finite-depth cylindrical capillary--gravity oscillations about contact-angle-dependent curved Young--Laplace equilibria do not appear to be available in the literature.
Results outside those benchmarked parameter ranges should accordingly be interpreted as predictions of the present surface-operator model.

\begin{table}
  \begin{center}
  \small
  \setlength{\tabcolsep}{4pt}
  \begin{tabular}{cccccccc}
    \toprule
    & & \multicolumn{3}{c}{$\lambda_{11}$} & \multicolumn{3}{c}{$\lambda_{12}$} \\
    \cmidrule(lr){3-5} \cmidrule(lr){6-8}
    $\theta_c$ & $N$ & present & benchmark & Rel. error & present & benchmark & Rel. error \\
    \midrule
    $15^\circ$  &  8 & 3.023 & 3.006 & $5.7\times10^{-3}$ &  51.047 &  50.100 & $1.89\times10^{-2}$ \\
    $30^\circ$  & 12 & 3.640 & 3.638 & $6.5\times10^{-4}$ &  76.181 &  76.670 & $6.37\times10^{-3}$ \\
    $45^\circ$  & 16 & 4.315 & 4.332 & $3.9\times10^{-3}$ & 104.646 & 104.600 & $4.40\times10^{-4}$ \\
    $60^\circ$  & 20 & 4.955 & 4.952 & $5.9\times10^{-4}$ & 129.186 & 129.400 & $1.65\times10^{-3}$ \\
    $75^\circ$  & 20 & 5.508 & 5.505 & $5.0\times10^{-4}$ & 145.651 & 145.900 & $1.70\times10^{-3}$ \\
    $90^\circ$  & 20 & 5.935 & 5.932 & $5.3\times10^{-4}$ & 151.535 & 151.800 & $1.74\times10^{-3}$ \\
    $105^\circ$ & 20 & 6.212 & 6.209 & $5.2\times10^{-4}$ & 146.430 & 146.700 & $1.84\times10^{-3}$ \\
    $120^\circ$ & 20 & 6.330 & 6.328 & $2.8\times10^{-4}$ & 132.070 & 133.100 & $7.74\times10^{-3}$ \\
    $135^\circ$ & 16 & 6.300 & 6.309 & $1.4\times10^{-3}$ & 112.313 & 115.400 & $2.67\times10^{-2}$ \\
    $150^\circ$ & 12 & 6.176 & 6.203 & $4.4\times10^{-3}$ &  93.492 &  98.940 & $5.51\times10^{-2}$ \\
    $165^\circ$ &  8 & 6.071 & 6.089 & $3.0\times10^{-3}$ &  83.391 &  87.500 & $4.70\times10^{-2}$ \\
    \bottomrule
  \end{tabular}
  \caption{Comparison for the first two eigenvalues, $\lambda_{11}$ and $\lambda_{12}$, with the benchmark values reported by \citet{kopachevskii1973} and reproduced in \citet{myshkis1987}, at $\Bo=0$, $h/R=1$ and $m=1$. The truncation order $N$ is adapted to the meniscus steepness.}
  \label{tab:myshkis}
  \end{center}
\end{table}

\subsection{Convergence and conditioning}

Table~\ref{tab:conv} reports $\lambda_{11}$ for $\theta_c=45^\circ$ and $\Bo=0$ as the truncation order $N$ increases.

\begin{table}
  \begin{center}
  \begin{tabular}{ccccc}
    \toprule
    $N$ & $\mathrm{cond}(\mathbf{A})$ & $n_{\mathrm{eff}}$ & $\lambda_{11}$ & Error vs. Myshkis \\
    \midrule
     4 & $2.2\times10^{1}$ &  4 & 4.316 & $3.6\times10^{-3}$ \\
     6 & $1.4\times10^{2}$ &  6 & 4.315 & $3.9\times10^{-3}$ \\
     8 & $1.1\times10^{3}$ &  8 & 4.315 & $3.9\times10^{-3}$ \\
    10 & $1.0\times10^{4}$ & 10 & 4.315 & $3.9\times10^{-3}$ \\
    12 & $9.5\times10^{4}$ & 12 & 4.315 & $3.9\times10^{-3}$ \\
    16 & $9.3\times10^{6}$ & 16 & 4.315 & $3.9\times10^{-3}$ \\
    20 & $9.5\times10^{8}$ & 20 & 4.315 & $3.9\times10^{-3}$ \\
    \bottomrule
  \end{tabular}
  \caption{Convergence of $\lambda_{11}$ with truncation order $N$ ($\theta_c=45^\circ$, $\Bo=0$, $m=1$, $h/R=1$). Here $\mathrm{cond}(\mathbf{A})$ is the condition number of the surface-coupling matrix and $n_{\mathrm{eff}}$ is the effective numerical rank at tolerance $\sigma_1/10^{12}$, where $\sigma_1$ denotes the largest singular value of $\mathbf{A}$.}
  \label{tab:conv}
  \end{center}
\end{table}

The leading eigenvalue is stable to four significant digits by $N=8$, whereas $\mathrm{cond}(\mathbf{A})$ rises from $2.2\times10^{1}$ at $N=4$ to $9.5\times10^{8}$ at $N=20$. The practical limitation is therefore not lack of spectral convergence but deterioration of conditioning in the trace solve. In the computations reported below, $N$ is taken as the smallest truncation for which the eigenvalues of interest are unchanged to the displayed accuracy under further increase of $N$; for steeper menisci this criterion is met at smaller $N$ because $\mathbf{A}$ becomes ill-conditioned more rapidly.

\subsection{Breakdown of the flat-interface approximation}
\label{ssec:flat-breakdown}

\begin{figure}[htbp]
  \centerline{\includegraphics[width=\textwidth,height=0.42\textheight,keepaspectratio]{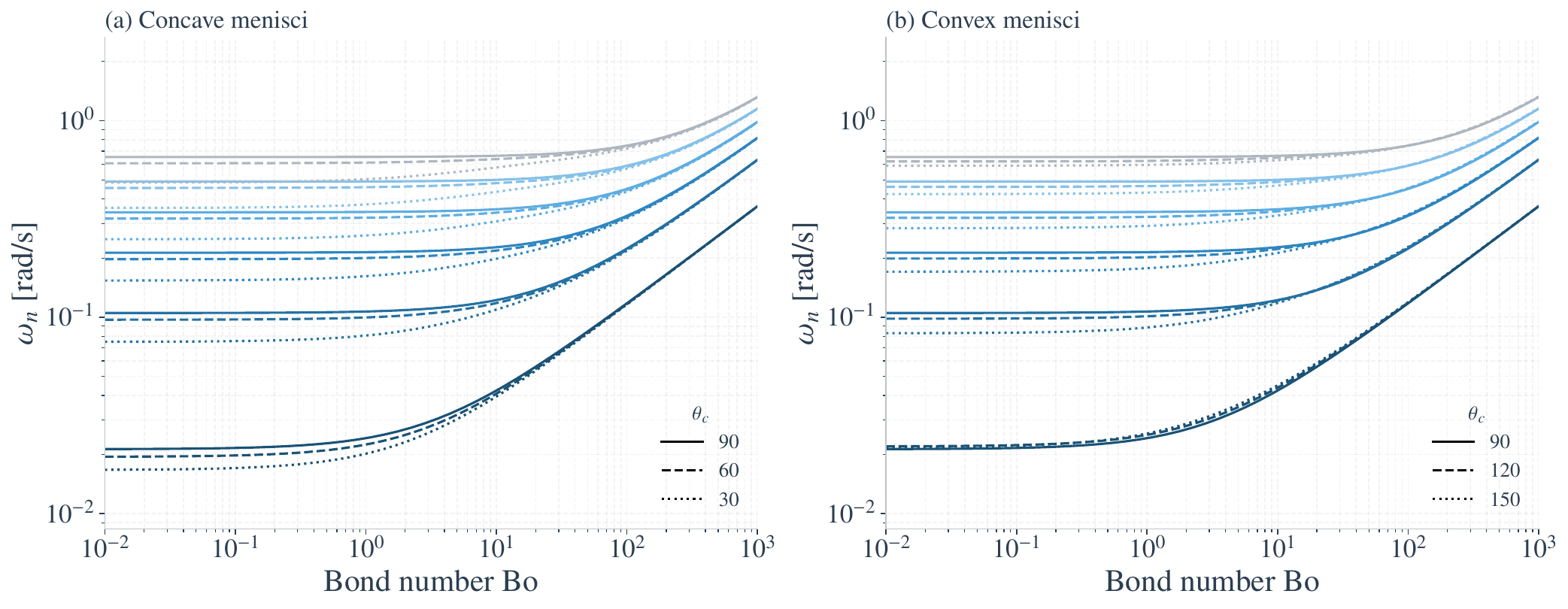}}
  \caption{First few sloshing frequencies $\omega_n$ in the fixed azimuthal sector $m=1$ as functions of the Bond number for representative concave and convex menisci at fixed fill depth $h/R=2$. In each panel, the solid curves correspond to the flat reference $\theta_c=90^\circ$, while the dashed and dotted curves correspond to the indicated contact angles. Each colour denotes a different radial branch. Deviations from the flat-interface spectrum are strongest at low Bond number. Concave menisci shift the branches downward, whereas convex menisci shift the fundamental branch upward, while the higher-order branches are shifted downward.}
  \label{fig:meniscus_shift}
\end{figure}

The discussion now turns to the predicted physical consequences of equilibrium curvature. Figure~\ref{fig:meniscus_shift} presents representative effects of the Bond number on the frequencies of the various modes for a fixed azimuthal mode, for wetting, neutral and non-wetting cases. Two robust trends emerge. First, the sign of the frequency correction changes across $\theta_c=90^\circ$: concave menisci ($\theta_c<90^\circ$) soften the fundamental mode, whereas convex menisci ($\theta_c>90^\circ$) stiffen the fundamental mode while softening the higher-order modes. Second, the correction becomes appreciable only when the Bond number is small enough for capillarity to control the equilibrium shape. By contrast, when gravity becomes dominant, the effect of curvature vanishes and the spectrum collapses to the flat-interface case.

\begin{figure}[htbp]
  \centerline{\includegraphics[width=\textwidth,height=0.42\textheight,keepaspectratio]{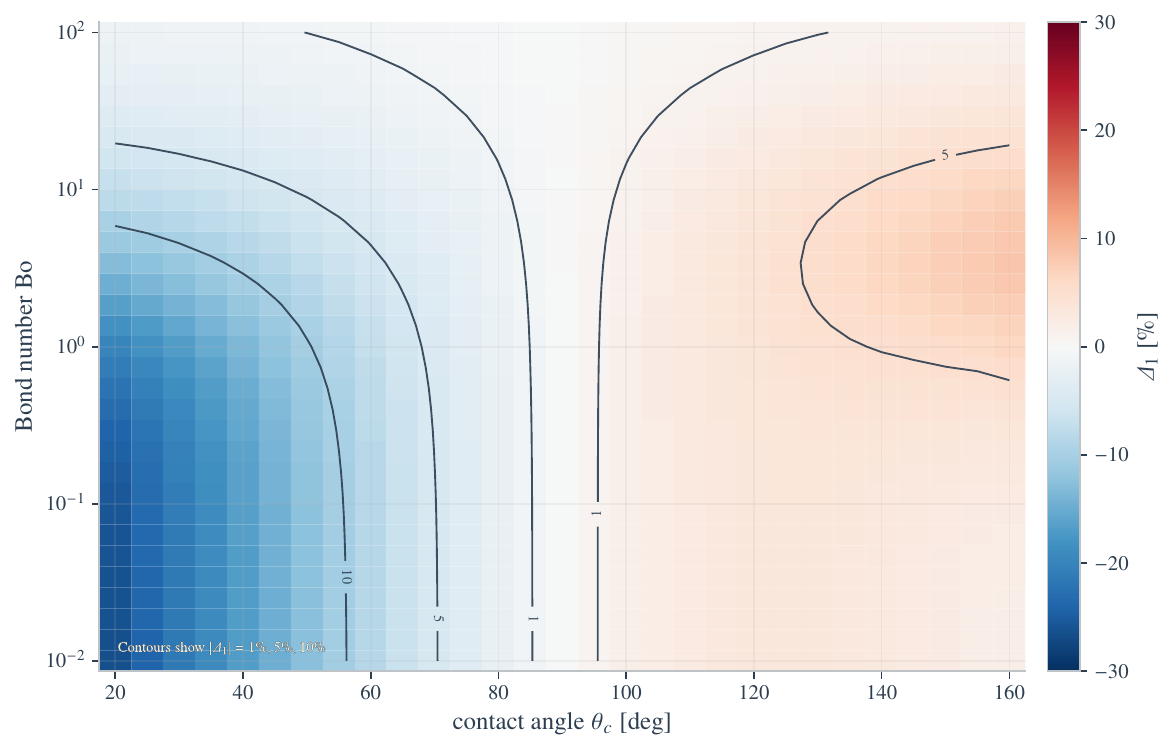}}
  \caption{Signed percentage change $\Delta_1$ in the fundamental frequency for the fixed azimuthal sector $m=1$ over the $(\theta_c,\Bo)$ plane, measured relative to the flat-interface value at the same Bond number. Contours at $|\Delta_1|=1\%$, $5\%$ and $10\%$ mark the progressive breakdown of the flat-interface approximation.}
  \label{fig:regime_map}
\end{figure}

A global view of these trends is provided by figure~\ref{fig:regime_map}, which maps the signed percentage change in the fundamental frequency,

\[
  \Delta_1 := 100\,
  \frac{\omega_{m1}(\Bo,\theta_c)-\omega_{m1}^{\mathrm{flat}}(\Bo)}
       {\omega_{m1}^{\mathrm{flat}}(\Bo)},
\]

over the $(\theta_c,\Bo)$ plane. Contours at $|\Delta_1|=1\%$, $5\%$ and $10\%$ then provide a direct criterion for the breakdown of the flat-interface approximation.

The regime map reveals that the flat theory remains accurate near terrestrial conditions. For $\Bo\gtrsim 10$ the correction stays below roughly $1\%$ unless the contact angle is very far from $90^\circ$. Around $\Bo=1$ the correction already reaches several per cent for representative wetting and non-wetting cases, and for $\Bo\lesssim 0.1$ it becomes an order-one effect on the scale of linear sloshing frequencies. The physically relevant transition is therefore not simply ``small Bond number'' in isolation, but the joint regime of low Bond number and contact angles sufficiently far from $\pi/2$ that the equilibrium meniscus acquires substantial curvature.

A further feature of the map is its asymmetry with respect to $\theta_c=90^\circ$: contact angles equally far from $\pi/2$ do not produce equal and opposite corrections. This asymmetry is already present in the benchmark data of table~\ref{tab:myshkis} and is therefore a property of the finite-depth cylindrical problem itself, not of the present discretisation. Its origin is analysed in \S\,\ref{ssec:operator}.

\subsection{Operator comparison and complementary-angle asymmetry}
\label{ssec:operator}

\begin{figure}[htbp]
  \centerline{\includegraphics[width=\textwidth,height=0.42\textheight,keepaspectratio]{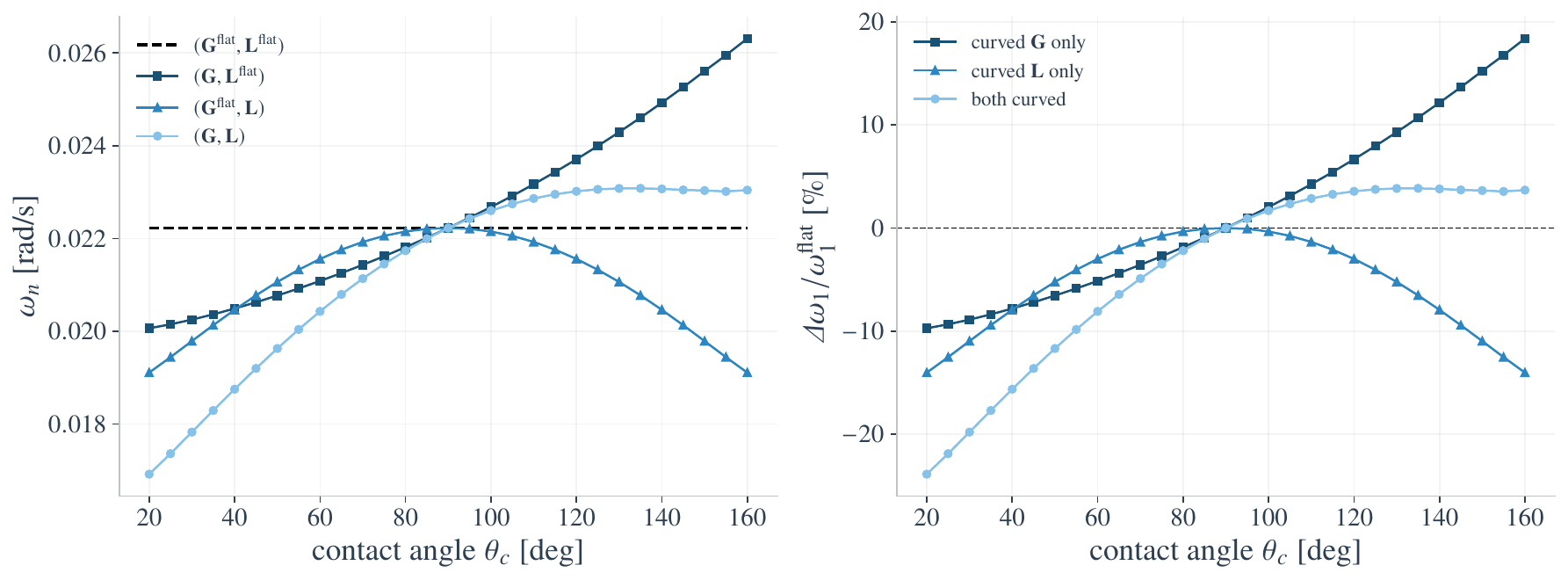}}
  \caption{Operator-split analysis at $\Bo=0.3$, $m=1$, $h/R=2$. Left: fundamental frequency for the four operator combinations. Right: percentage shift relative to the flat--flat reference. The fully curved result is not reproduced by either hybrid problem alone, confirming that curvature modifies both $\mathbf{G}$ and $\mathbf{L}$.}
  \label{fig:operator_split}
\end{figure}

To separate the respective contributions of the kinetic and capillary operators, figure~\ref{fig:operator_split} compares the fundamental frequency for four operator combinations: a flat--flat reference, a curved-$\mathbf{G}$/flat-$\mathbf{L}$ hybrid, a flat-$\mathbf{G}$/curved-$\mathbf{L}$ hybrid, and the fully curved problem. The fully curved frequency is not, in general, recovered by either hybrid problem in isolation, showing that curvature modifies both operators and the eigenvector that balances them. In some parameter ranges the curved Dirichlet--Neumann operator accounts for most of the shift. In others the capillary correction is comparable, and in the strongly curved regime both contributions are essential.

This operator-level view also explains the complementary-angle asymmetry noted in \S\,\ref{ssec:flat-breakdown}. At fixed volume the equilibrium menisci at $\theta_c$ and $\pi-\theta_c$ satisfy

\[
  H_{\pi-\theta_c}(r)=2h-H_{\theta_c}(r),
  \qquad
  H'_{\pi-\theta_c}(r)=-H'_{\theta_c}(r).
\]

The capillary form \eqref{eq:Lpq} depends only on $H_0'^{\,2}$ and is therefore invariant under this reflection. The mapped Laplace operator \eqref{eq:laplace-mapped}, however, is not: with $J=H_0$ and $\alpha=H_0'$, the replacement $J\mapsto 2h-J$, $\alpha\mapsto-\alpha$ alters the coefficients of the $\pp_{\zeta\zeta}\phi$, $\pp_{r\zeta}\phi$ and $\pp_\zeta\phi$ terms. The asymmetry is therefore a genuine kinetic effect carried by $\mathbf{G}$, consistent with the same benchmark values reported by \citet{kopachevskii1973} and reproduced in \citet{myshkis1987}. Figure~\ref{fig:operator_split} (right panel) confirms this decomposition: the capillary-only curve is nearly symmetric about $90^\circ$, whereas the kinetic-only curve is monotonic, and the full correction inherits its asymmetry from $\mathbf{G}$.

\begin{figure}[htbp]
  \centerline{\includegraphics[width=\textwidth,height=0.42\textheight,keepaspectratio]{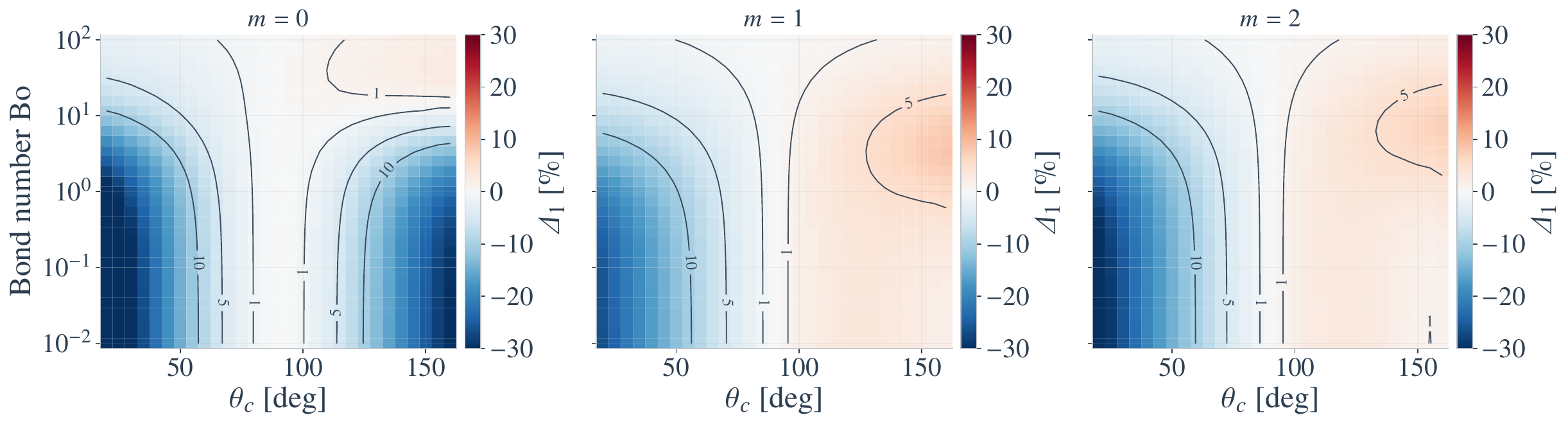}}
  \caption{Regime maps of the signed percentage frequency shift $\Delta_1$ for azimuthal modes $m=0$, $1$ and $2$ ($h/R=2$). All three modes display the same qualitative structure: negligible correction for $\Bo\gtrsim 10$, progressive breakdown of the flat-interface approximation for $\Bo\lesssim 1$, and wetting/non-wetting asymmetry.}
  \label{fig:regime_multimode}
\end{figure}

Finally, figure~\ref{fig:regime_multimode} extends the regime-map analysis to the axisymmetric ($m=0$) and second azimuthal ($m=2$) modes, confirming that the phenomenology is not specific to $m=1$: all three modes exhibit qualitatively similar patterns of curvature-induced frequency shift, with the transition from negligible to order-one corrections occurring in the same region of the $(\theta_c,\Bo)$ plane.

\subsection{Curvature-induced deformation of low-order modes}

\begin{figure}[htbp]
  \centerline{\includegraphics[width=\textwidth,height=0.32\textheight,keepaspectratio]{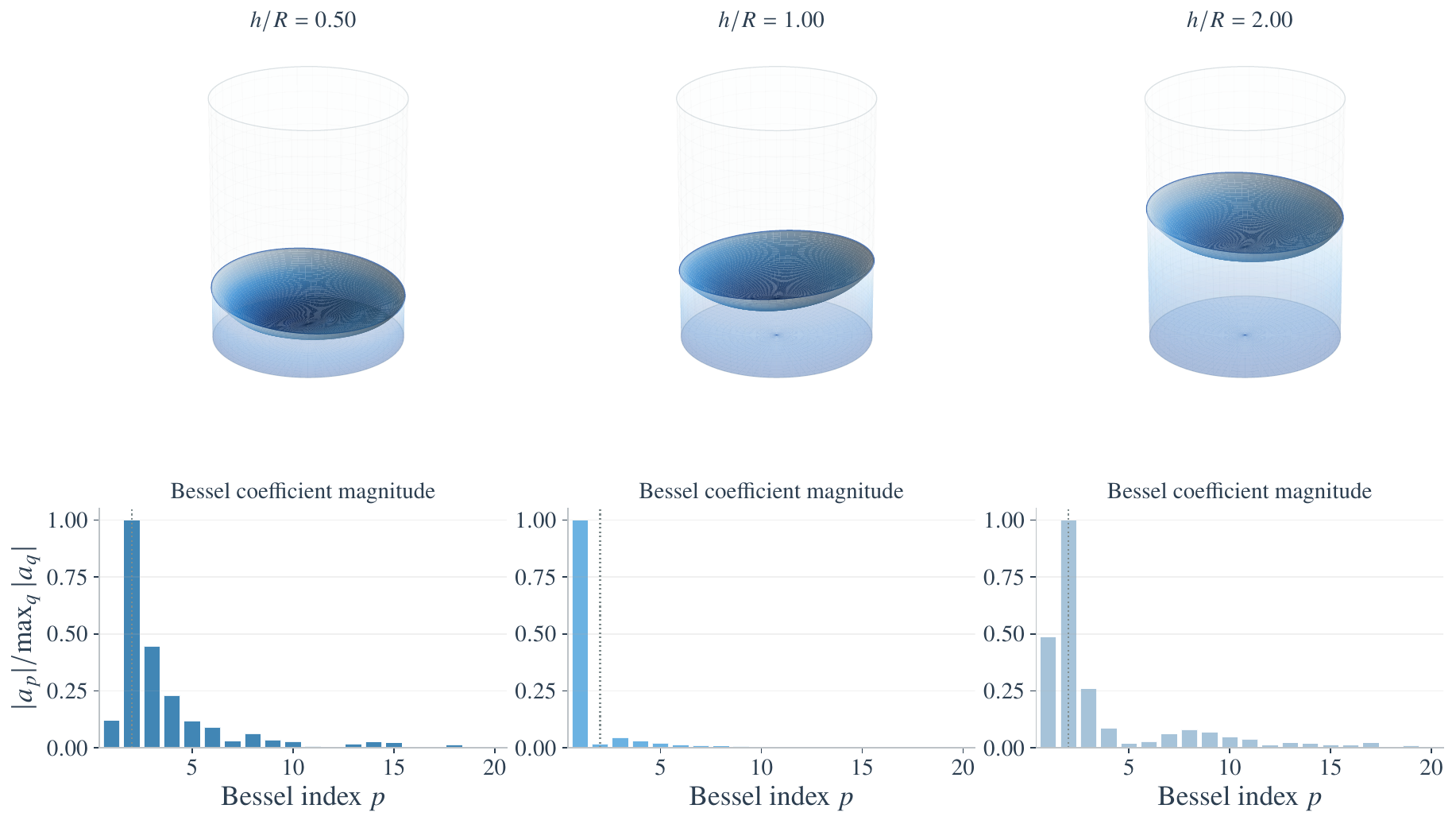}}
  \caption{Mode geometry and modal composition at fixed contact angle $\theta_c=20^\circ$, Bond number $\Bo=0.3$ and azimuthal sector $m=1$, shown for three representative fill levels $h/R=0.5$, $1.0$ and $2.0$. Upper row: three-dimensional view of the displaced free surface $H_0\pm\widehat{\eta}$ inside the cylindrical tank for the displayed eigenmode. Lower row: normalised Bessel coefficient magnitudes $|a_p|/\max_q |a_q|$, showing how the modal composition varies with fill level. The coefficient magnitude is most strongly concentrated in the $p=1$ component at $h/R=1$, whereas shallower and deeper fills display stronger redistribution across neighbouring radial orders.}
  \label{fig:fill_levels}
\end{figure}

For each eigenmode shown below, let

\[
  \widehat{\eta}(r,\theta)=\sum_p a_p\,\varphi_p(r)e^{im\theta}
\]

denote the reconstructed free-surface mode shape, with coefficients $a_p$ in the Bessel basis. Figure~\ref{fig:fill_levels} first shows how fill level alters the modal composition of a representative wetting case at fixed $\theta_c=20^\circ$, $\Bo=0.3$ and $m=1$. For the mode displayed, the normalised Bessel coefficient magnitude is most strongly concentrated in the $p=1$ component at $h/R=1$, whereas the $h/R=0.5$ and $h/R=2$ cases exhibit visibly broader participation of neighbouring radial orders. Fill level therefore modulates the strength of curvature-induced mixing, even though the complementary-angle asymmetry discussed above remains a kinetic effect carried by $\mathbf{G}$.

\begin{figure}[htbp]
  \centerline{\includegraphics[width=\textwidth,height=0.32\textheight,keepaspectratio]{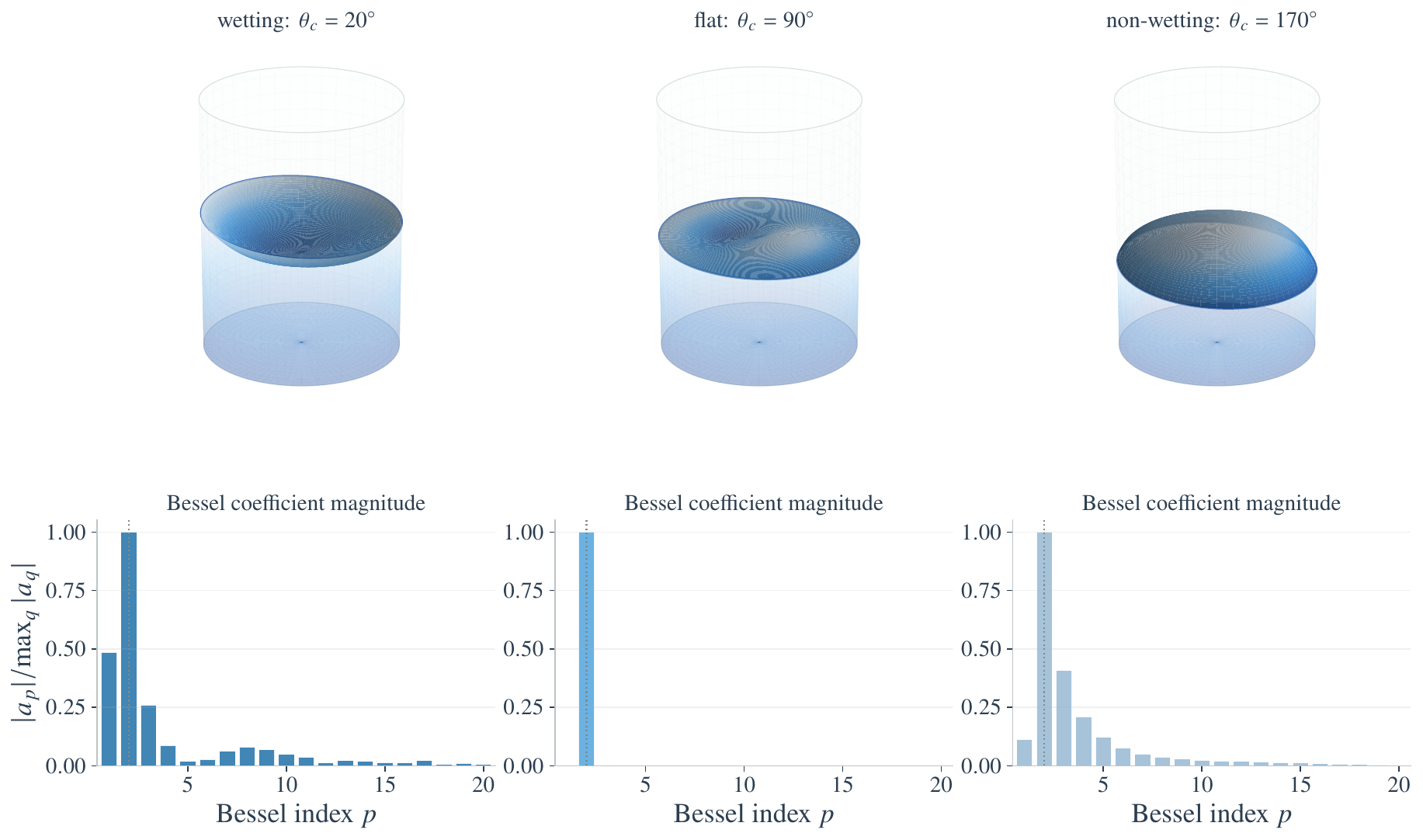}}
  \caption{Mode geometry and modal composition for the first mode in the fixed azimuthal sector $m=1$ at $\Bo=0.3$ and $h/R=2$, shown for three representative contact angles. Upper row: three-dimensional view of the displaced free surface $H_0\pm\widehat{\eta}$ inside the cylindrical tank, coloured by surface elevation. Lower row: normalised Bessel coefficient magnitudes $|a_p|/\max_q |a_q|$, showing how equilibrium curvature redistributes the mode across radial orders.}
  \label{fig:fundamental_mode}
\end{figure}

Figure~\ref{fig:fundamental_mode} then shows the first mode in the fixed azimuthal sector $m=1$ for one concave case, the flat case and one convex case at the same $(\Bo,h/R)$. The upper row displays the equilibrium interface together with the displaced surfaces $H_0\pm\widehat{\eta}$ inside the cylindrical tank. The lower row reports the normalised Bessel coefficient magnitudes $|a_p|/\max_q |a_q|$ of the corresponding eigenvector.

For the flat interface this mode is essentially a single Bessel profile and the eigenvector is concentrated in the leading radial component. Once the meniscus is curved this purity is lost: neighbouring radial orders acquire non-negligible weight and the free-surface shape can no longer be described by a single $J_1(\epsilon_{11}r/R)$ profile. This modal mixing is a direct manifestation of the off-diagonal entries in $\mathbf{G}$ and $\mathbf{L}$. Curvature therefore modifies not only the oscillation frequency but also the spatial structure of the mode.

\begin{figure}[htbp]
  \centerline{\includegraphics[width=\textwidth,height=0.32\textheight,keepaspectratio]{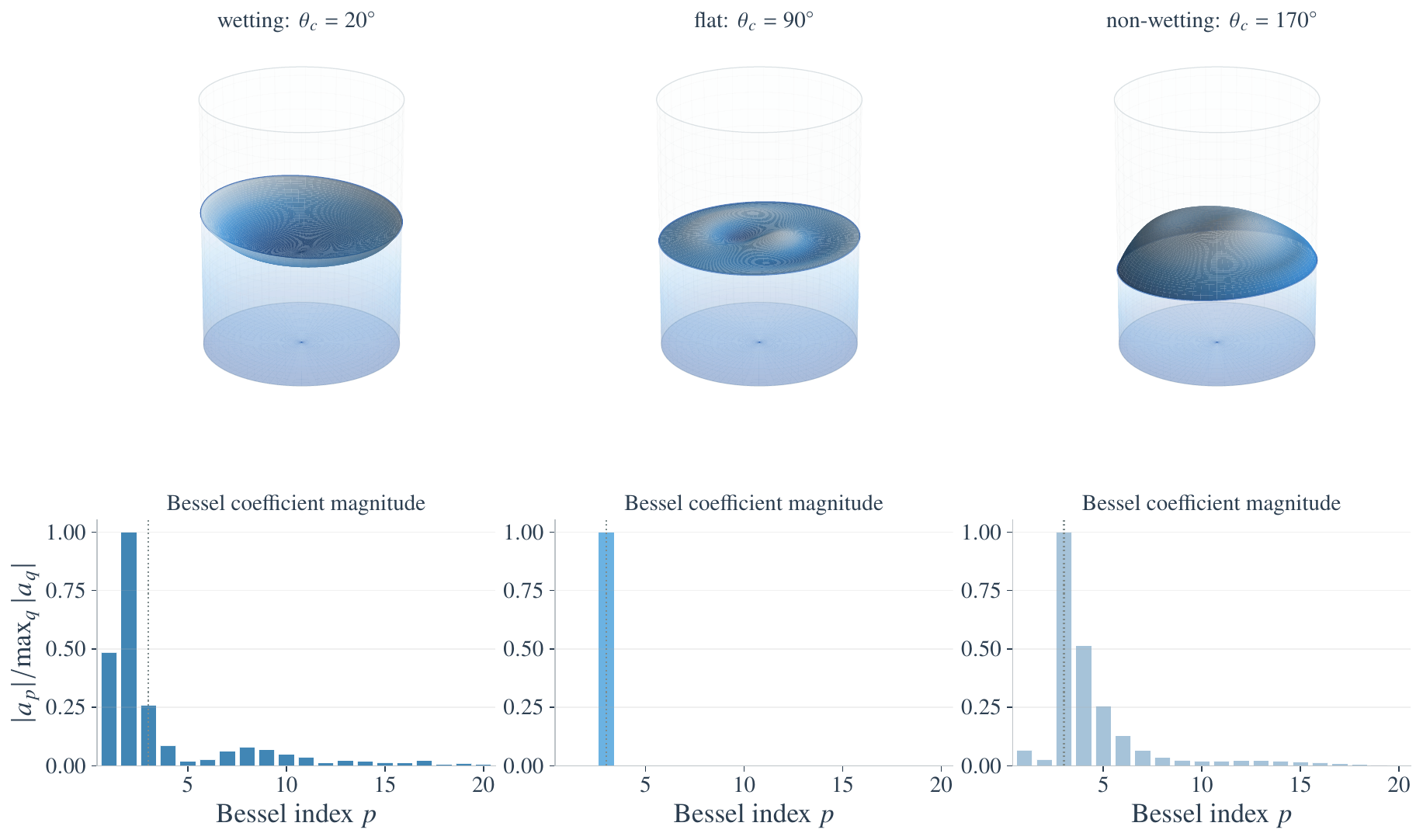}}
  \caption{Mode geometry and modal composition for the second mode in the fixed azimuthal sector $m=1$ at $\Bo=0.3$ and $h/R=2$, shown for three representative contact angles. Upper row: three-dimensional view of the displaced free surface $H_0\pm\widehat{\eta}$ inside the cylindrical tank, coloured by surface elevation. Lower row: normalised Bessel coefficient magnitudes $|a_p|/\max_q |a_q|$, showing how equilibrium curvature redistributes the mode across radial orders.}
  \label{fig:second_mode}
\end{figure}

Figure~\ref{fig:second_mode} then shows the second mode in the same azimuthal sector $m=1$. As in the fundamental branch, curvature spreads the second mode across neighbouring Bessel components, showing that mode mixing persists across successive radial branches rather than being confined to the lowest mode. Together with the regime maps and operator split, these results show that curvature changes both the frequencies and the modal structure of the low-order cylindrical spectrum.


\section{Conclusions}
\label{sec:conclusions}

A semi-analytical cylindrical surface-operator formulation has been developed for linear capillary--gravity sloshing about curved Young--Laplace menisci in microgravity. The formulation recovers the classical flat-interface spectrum exactly and remains in close agreement with the cylindrical benchmark values reported by \citet{kopachevskii1973} and reproduced in \citet{myshkis1987}. Its main contribution is to retain the Bessel structure of the tank while exposing the curved-meniscus eigenproblem as the interaction of a bulk inertial operator and a capillary restoring operator. Because the Dirichlet--Neumann operator and the linearised curvature operator remain explicit, curvature-induced frequency shifts can be decomposed into kinetic and capillary parts, and the observed wetting/non-wetting asymmetry can be attributed physically rather than read only from the final eigenvalues. The formulation therefore turns the classical cylindrical low-Bond spectrum into a form that is not only benchmarked, but also mechanically interpretable.

The calculations show that equilibrium curvature is negligible only when $\Bo\gg 1$. Once $\Bo\lesssim 1$, it becomes a leading-order dynamical effect: the low-order sloshing frequencies depart appreciably from the flat-interface prediction, the associated eigenmodes are redistributed across neighbouring radial Bessel components, and in strongly capillary regimes the shift is large enough to alter the dominant time scales of the liquid motion. This is not a minor geometric correction. Natural sloshing frequencies are precisely the quantities that enter reduced-order models used for guidance, navigation and control, resonance-avoidance assessments, and the interpretation of propellant-induced forces and torques \citep{simonini2024}. In the low-Bond regime, using the flat-interface spectrum can therefore misplace the relevant dynamical scales of the tank and lead to physically misleading predictions.

The operator decomposition also makes explicit what is not transparent from the classical cylindrical benchmark spectrum alone. The capillary contribution remains nearly symmetric with respect to complementary contact angles, whereas the wetting/non-wetting asymmetry is carried primarily by the Dirichlet--Neumann operator and is therefore predominantly kinetic. In that sense, the present formulation does not replace the classical cylindrical results collected by \citet{myshkis1987}; it explains them more sharply, by identifying which part of the model generates the observed asymmetry and frequency shift. Curved equilibrium menisci should thus be regarded as part of the leading-order model of cylindrical microgravity sloshing, not as a secondary refinement. For axisymmetric equilibrium menisci, the linearised problem decouples into azimuthal Fourier sectors, while genuinely non-axisymmetric equilibrium interfaces would require a two-dimensional surface discretisation together with a corresponding bulk representation. The present formulation therefore provides a benchmarked route beyond the flat-interface approximation and, more importantly, a sharper physical interpretation of the classical curved-meniscus spectrum.

\paragraph*{Acknowledgements}
The author used GitHub Copilot Chat (GPT-5.4, accessed via the GitHub Copilot extension in Visual Studio Code on 29 April 2026) to assist with language editing, rephrasing and restructuring parts of the manuscript text. All scientific content, equations, references, results and final wording were reviewed and approved by the author, who takes full responsibility for the manuscript.

\paragraph*{Funding}
This work was supported by the Italian National Recovery and Resilience Plan (PNRR) and by Thales Alenia Space Italia.

\paragraph*{Declaration of interests}
The author reports no conflict of interest.

\paragraph*{Data availability}
The data generated during this study and the custom scripts used to produce the reported results and figures are available from the corresponding author upon request.

\end{document}